# Understanding Formulation of Social Capital in Online Social Network Sites (SNS)

S.S.Phulari,[1], Dr.S.D.Khamitkar[2], N.K.Deshmukh[3], P.U.Bhalchandra[4], S.N.Lokhande[5] and A.R.Shinde[6]

[1,2,3,4,5] School of Computational Sciences,
Swami Ramanand Teerth Marathwada University, Nanded, MS, India, 431606

[6] Department of Computer Science,
N.S.B. College, Nanded, MS, India, 431601

## Abstract

Online communities are the gatherings of like-minded people, brought together in cyberspace by shared interests. The shared interest has hidden social capital aspects and can be of bridging or bonding type .Creating such communities is not a big challenge but sustaining member's participation is. This study examines the formation and maintenance of social capital in social network sites. In addition to assessing bonding and bridging social capital, we explore a dimension of social capital that assesses one's ability to stay connected with members of a previously inhabited community, which we call maintained social capital. Such dimension is enacted here in terms of Hypothesis.

## 1. Introduction

Social network are the graph that represents relationships between users. These can take any shape, including

- **Offline Social connections** (friends, clubs, groups, associations)
- **Online social connections** (Face book, MySpace, Live Journal, Orkut, LinkedIn)
- **Messaging** (IM, chat, address book)
- **Social book marking** (Digg, Delicious)
- **Content sharing** (Flickr, YouTube)

It is evident from the exponential growth of Internet users in recent year that, more and more of people's interactions are happening online. Hence, now days, social networks are commonly seen to be associated with Internet. Such networks must get studied with the type of social correlation the user community has. We found three sources of social correlation as defined below,

**Influence:** An individual performing an action can cause some contacts to do the same by providing information or by increasing the value of the action to them.

**Homophily:** Similar individuals are more likely to be friends. For example two geologists are more likely to know each other than two random people.

**Environment:** Influence from external elements. Friends are more likely to live in the same area, thus attend and take pictures of similar events, and tag them with similar tag. This focus on a particular action like buying a product, joining a community, publishing in a conference, using a particular tag, getting premier membership, etc

Social network analysis goes back to almost half a century of work in sociology and social psychology. *Milgram* in 1967 attempted to define it in concise form and stated that two random people were connected on average by a path of six acquaintances. This is called as six degrees of separation theorem. Similarly, *Zachary* in 1972 discussed social relationships and rivalries in a University Karate Club (via graph min cuts). In recent years, as the world is migrating to global village or hyper world, large number of social networks found activated, with thousands of users associated? For theorist computer scientists, this exponential growth or the resulting data is a goldmine for giving a new perspective via sociological studies, data mining studies, pattern matching studies, etc. This data is of massive amounts, as hundreds of millions of users are there. This data has a study phenomenon at different scales, for example interaction of people in focused groups of different sizes, overall structure of the network including regional and outside-region support, etc. This data has ability to measure, record, and track individual activities at the finest resolution also. For example, user befriending another, user buying a DVD, user tagging a photo, etc.

This study is important today in order to find out behavior pattern of users connected via social networks. This helps us to predict the dynamics of the network system also. At the same time, this study help us to talk about new norms of behavior, technology, influences or idea, we can discover after analyzing data by questioning and analyzing how similar is the behavior of connected users?, how similar the structure of services and technologies social network sites have? .We can also device key indictors of social networks from such study.

A popular Social Network researcher, Wellman [1] argued that the concept of "social networks" is difficult to define. Usually a social network is defined as relations among people who deem other network members to be important or relevant to them in some way [2]. Using media to develop and maintain, social networks is established in practice. Now days, this concept is popularly implemented in term of Internet sites. These websites allow participants to construct a public or semi-public profile within the system and formally articulate their relationship to other users in a way that is visible to anyone who can access their profile [3, 4] We feel that this definition does not specify the closeness of any given connection, but only that participants are linked in some fashion.

Beside, being just a gatherings of like-minded people, brought together in cyberspace by shared interests linkage, Social Network Sites (SNSs) such as such as MySpace allow individuals to present themselves, articulate their social networks, and establish or maintain connections with others. These sites can be oriented towards work-related contexts like LinkedIn.com, romantic relationship initiation like Friendster.com, and for connecting those with shared interests such as music or politics, or the college student population like the Facebook.com.

The SNS operational style can be online or offline. The offline style has no face to face communications. The online social network application enables its users to present themselves in an online profile, accumulate "friends" who can post comments on each other's pages, and view each other's profiles. For example, Face book members can also join virtual groups based on common interests, see what classes they have in common, and learn each others' hobbies, interests, musical tastes, and romantic relationship status through the profiles. Online SNSs support both the maintenance of existing social ties and the formation of new connections. Early research on online communities assumed that individuals using these systems would be connecting with others outside their pre-existing social group or location, liberating them to form communities around shared interests, as opposed to shared geography [2, 5].This could be interesting to predict about social capital. A hallmark of this early research is the presumption that when online and offline social networks overlapped, the directionality was *online to offline*-online connections resulted in face-to-face meetings. Although early work on SNS addressed many interesting findings, we find no traces of social capital. This is because there is little empirical research that addresses whether members use SNSs to maintain existing ties or to form new ones, the social capital implications of these services is unknown.

The SNS like Face book constitutes a rich site for researchers interested in the affordances of social networks due to its heavy usage patterns and technological capacities that bridge on line and offline connections. Research suggests that Face book users engage in "searching" for people with whom they have an offline connection more than they "browse" for complete strangers to meet [4]. We believe that this approach in Face book represents an understudied offline to online trend where it originally, primarily served a geographically-bound community (the campus). However, as social capital researchers we have another perspective to look into – findings of social capital as place-based community facilitate the generation of social capital.

We are optimistic to get many social capitals and if a Regression analysis is conducted on them then it will definitely suggest a strong association between use of SNS and the type of social capital, with the strongest relationship being to bridging social capital. In addition we could also find hidden aspects to interact with the measures of psychological well-being. This discovery may help in suggesting that social capital and SNS can provide greater benefits for users experiencing low self-esteem and low life satisfaction. This is because a depressed person has more tendencies to express feeling than a normal person. [6].Such study brings a requirement to statistically, analytically process SNS data. However a social network will definitely look different depending upon how one measures it, like, counting the number of interactions between members (say "a") or rating the closeness of relationships (say "b") or "a versus b", etc .

## 2. An Overview of Face book

Created in 2004, by 2007 Face book was reported to have more than 21 million registered members generating 1.6 billion page views each day The site is tightly integrated into the daily media practices of its users: The typical user spends about 20 minutes a day on the site, and two-thirds of users log in at least once a day [7]. Capitalizing on its success it created separate versions for high school students, communities for commercial organizations.

Face book s widely used for SNA due to its openness and large customer base. Much of the existing academic research on Face book has focused on identity presentation and privacy concerns [8]. Looking at the amount of information Face book participants provide about themselves, the relatively open nature of the information, and the lack of privacy controls enacted by the users, argue that users may be putting themselves at risk both offline (e.g., stalking) and online (e.g., identify theft). Other recent Face book research examines student perceptions of instructor presence and self-disclosure [9] temporal patterns of use [10] and the relationship between profile structure and friendship articulation [4].

The literature survey [7] shows that Face book is used more by college-age students and was significantly associated with measures of social capital. We use Face book as a research context in order to determine whether knowledge about social capital patterns can be generated by enacting some theorems or hypothesis.

## 3. Defining Social Capital

Social capital is an elastic term. It broadly refers to the resources accumulated through the relationships among people. It has variety of definitions [11]. It is often conceived as a cause and an effect for social networking. For individuals, social capital allows a person to draw on resources from other members of the networks to which he or she belongs. These resources can take the form of useful information, personal relationships, or the capacity to organize groups [12].This also give capabilities to access to individuals outside one's close circle and help to get non-redundant information, resulting in benefits such as employment connections [13].

Social capital has been linked to a variety of positive social outcomes, such as better public health, lower crime rates, and more efficient financial markets [13].When social capital declines, a community experiences increased social disorder, reduced participation in civic activities, and potentially more distrust among community members. Greater social capital increases commitment to a community and the ability to mobilize collective actions, among other benefits. Social capital may also be used for negative purposes, but in general social capital is seen as a positive effect of interaction among participants in a social network [14]. The Internet has been linked both to increases and decreases in social capital [1]. We can probe into social capital in two ways,

**Bridging (weak ties):** loose connections between individuals which may provide useful information or new perspectives for one another

**Bonding (strong ties):** strong connection between individuals. Usually emotionally close relationships, such as family and close friends.

Recently, researchers have emphasized the importance of Internet-based linkages for the formation of weak ties, which serve as the foundation of bridging social capital. Because online relationships may be supported by technologies like distribution lists, photo directories, and search capabilities [15], it is possible that new forms of social capital and relationship building will occur in online social network sites. Bridging social capital might be augmented by such sites, which support loose social ties, allowing users to create and maintain larger, diffuse networks of relationships from which they could potentially draw resources [10].

## 4. Understanding Formulation of Social Capital

After briefly describing, to the extant of literature [9, 11, 16, 17, and 18] regarding the forms of social capital and the impact of Internet, we introduce some basic hypothesis for social capital which speaks to the ability to maintain valuable connections as one progress through life changes.

4.1. Hypothesis 1: Use of SNS will be proportional to user's perceived bonding social capital.

**Explanation:** Bonding social capital reflects strong ties with family and close friends. Day by day, internet is maturing and providing new means, connections to connect and come closer. These new connections may result increase in social capital as users might be in a position to provide emotional support whenever needed. Thus, as the tendency to share interests or regional goals increases, user starts bonding the social capitals and use of SNS will increase subsequently. This is the reason why terrorists frequently use SNS like Orkut and Face Book. Online users are more likely to have a larger network of close ties than non-Internet users, and Internet users are more likely than non-users to receive help from core network members. This proves our hypothesis.

### 4.2. Hypothesis 2: Bridging social capital is largely affected by the psychological status of users across SNS

**Explanation:** Usually SNS help an individual who have difficulties .Consider a research scholar having less economical affordability. For completion of the project he or she has undertaken, he or she will definitely need availability of full size citation papers. Considering the time constraint associated with the research project and because of sever economical backwardness, he or she may become panic and get attached to contemporary research platforms via SNS where he or she could find some of research literature by bridging social capital. We will not observe bonding social capital because the connection established is temporary and does not reflect long term, core and sturdy relationships. If his or her economical conditions were good or there was no narrow time constraint with the project, then the person would not became panic and would never had accessed SNS / internet. Thus SNS may be useful for individuals who have difficulties or individuals with low psychological levels .However bridging social capital will vary depending on the degree of a person's self esteem or depending on the degree of a person's satisfaction with life.

### 4.3. Hypothesis 3: Change in user's life style adversely affects social capital

**Explanation:** Social networks change over time as relationships are formed or abandoned. Particularly significant changes in social networks may affect one's social capital, as when a person moves from a geographic location in which his network was formed and thus loses access to those social resources. The possible causes of decreased social capital can be many including increase in families moving for job reasons, natural disasters, migrations, or even discrete internet connectivity which results increasing tendency to rely on emails for long distance communications. The time zone, working timings also affect the social capital .The connectivity between parents and children pursuing their education abroad have affected social capital as timing zones are different and it is highly impossible to remain online. We found that services like email and instant messaging help college students remain close to their high school friends after they leave college for jobs. This is known as "Fiendsickness". Young adults moving from high schools often leave friends from high school with whom they may have established rich networks; completely abandoning these high school networks would mean a loss of social capital.

### 4.4. Hypothesis 4: A bridging social capital always maintain the social capital

**Explanation:** Literature review suggests that the weak ties provide more benefit when the weak tie is not associated with stronger ties, as may be the case for maintained high school relationships. This is because the high school relations are friendship relations and have no much emotional, business gravities. However this is subject to psychological aspects, which rarely play a role. For example, a child of divorced parents or a child of working parents needs a bonding relationship with his friend. Such ties are only maintained in high school relationships even after leaving schools .We call this concept "maintained social capital."

## 5. Conclusions

The findings of this exploratory study on social capital suggest that research efforts are required in the area of user behavior and SNS modeling, with a special attention to the definition social capital. We have elaborated three measures of social capital including bridging, bonding, maintained social capital and analyzed user's behavior with respect to them. It is evident from our work that internet is linked to both increase and decrease in social capital. After briefly describing to the extant of literature on these three forms of social capital and the Internet, we introduced an additional dimension of social capital that speaks to the ability to maintain valuable connections as one progress through life changes. The conclusions put here are mainly for most popular SNS-Face Book, as much of the existing academic research has been focused on it. This could be due to its relatively open nature. Basic theorems are also mentioned to support our findings.

**References**
[1] Wellman, B., Haase A., Witte J. and Hampton K "Does the Internet Increase, Decrease, or Supplement Social Capital? , appeared in Social Networks, Participation and Community Commitment, American Behavioral Scientist, issue 45, November 2001.
[2] Wellman B, Computer Networks as Social Networks, Science press,2001, pp 231-234.
[3] Danah Michele Boyd, Friendster and publicly articulated social networking, extended abstracts on Human factors in computing systems, April 2004, pp 24-29, Vienna, Austria [DOI-10.1145/985921.986043]
[4] Ellison N., Lampe C. and Steinfield C.,Spatially Bounded Online Social Networks and Social Capital: The Role of Face book, International Conference on Communication Association, Dresden, 2006.

[5] Adam N. Joinson,Looking at, looking up or keeping up with people?: motives and use of face book, Proceeding of the twenty-sixth annual SIGCHI conference on Human factors in computing systems, 2008, pp 05-10, Florence, Italy

[6] Utz S,the Development of Friendships in Virtual Worlds, Social Information Processing in MUDs: Journal of Online Behavior, 2000.

[7] Jennefer Hart , Charlene Ridley , Faisal Taher , Corina Sas , Alan Dix,Exploring the face book experience: a new approach to usability, Proceedings of the 5th Nordic conference on Human-computer interaction: building bridges, October 2008, Lund, Sweden

[8] Cliff A. C. Lampe, A familiar face (book): profile elements as signals in an online social network, Conference on Human Factors in Computing Systems proceedings of the SIGCHI conference on Human factors in computing systems, San Jose, California, USA

[9] Stutzman, F, An Evaluation of Identity-Sharing Behavior in Social Network Communities, appeared in iDMAa and IMS Code Conference, Oxford, 2005.

[10] Sunnafrank M, Predicted outcome value during initial interactions: A reformulation of uncertainty reduction theory, (1986), Journal of Human Communication Research, 13. pp 3-33.

[11] Joan DiMicco , David R. Millen , Werner Geyer , Casey Dugan , Beth Brownholtz , Michael Muller,Motivations for social networking at work, Proceedings of the ACM 2008 conference on Computer supported cooperative work, November 2008, San Diego, CA, USA

[12] John M. Carroll , Mary Beth Rosson , Theorizing mobility in community networks, , International Journal of Human-Computer Studies, v.66 n.12, , Dec. 2008 ,pp.944-962 .

[13] Jenny Preece , Diane Maloney-Krichmar, Focusing on sociability and usability, Online communities: The human-computer interaction handbook: fundamentals, evolving technologies and emerging applications, Lawrence Erlbaum Associates, Inc., Mahwah, NJ, 2002

[14] Adam N. Joinson,Looking at, looking up or keeping up with people?: motives and use of facebook, Proceeding of the twenty-sixth annual SIGCHI conference on Human factors in computing systems, April 2008, Florence, Italy

[15] Leysia Palen , Sarah Vieweg,the emergence of online wide scale interaction in unexpected events: assistance, alliance & retreat, Proceedings of the ACM 2008 conference on Computer supported cooperative work, November 2008, San Diego, CA, USA

[16] Charles Steinfield , Joan M. DiMicco , Nicole B. Ellison , Cliff Lampe,Bowling online: social networking and social capital within the organization, Proceedings of the fourth international conference on Communities and technologies, June 2009, University Park, PA, USA

[17] Ellison, N., Lampe, C. and Steinfield, C, Spatially Bounded Online Social Networks and Social Capital: The Role of Face book. Proceedings of Journal of International Communication Association, Dresden, 2006.

[18] Brain Knudsen, Richard Florida and Denise Rousseau , Dissertation Report on Bridging and Bonding – A multi Dimensional Approach to Regional Social Capital ,the Martin Prosperity Institute , University of Toronto , 2007

**S.S.Phulari:** He is a research student working on social area networking under the supervision of Dr.S. D. Khamitkar. He has completed his M.Phil and has 2+ papers in international conferences/ journals

**Dr.S.D.Khamitkar :** He has 14+ years experience in teaching and research. He is supervising ten research students currently and has 8+ papers in international conferences and journals

**N.K.Deshmukh,S.N.Lokhande and P.U.Bhalchandra :** These are faculties at University Department and have 5+ research papers in international conferences and journals .

**A.R.Shinde :** He is a faculty at one of the affiliated college and has ten years teaching experience. He is currently planning for research work and actively involved in understanding research process.